\documentclass[12pt]{article}
\usepackage{amsmath,amsfonts,bbold,esint,hyperref,graphicx,color}

\newcommand{\be}{\begin{equation}}
\newcommand{\ee}{\end{equation}}
\newcommand{\bea}{\begin{eqnarray}}
\newcommand{\eea}{\end{eqnarray}}
\newcommand{\beas}{\begin{eqnarray*}}
\newcommand{\eeas}{\end{eqnarray*}}

\newcommand{\Hext}{\overset{\leftrightarrow}{H}\rlap{\phantom{H}}}
\def\identity{{\mathbb{1}}}

\begin{document}
\begin{titlepage}

\begin{center}

{\Large Endpoint contributions to excited-state modular Hamiltonians}

\vspace{12mm}

\renewcommand\thefootnote{\mbox{$\fnsymbol{footnote}$}}
Daniel Kabat${}^{1}$\footnote{daniel.kabat@lehman.cuny.edu},
Gilad Lifschytz${}^{2}$\footnote{giladl@research.haifa.ac.il},
Phuc Nguyen${}^{1,2}$\footnote{phuc.nguyen@lehman.cuny.edu},
Debajyoti Sarkar${}^{3}$\footnote{dsarkar@iiti.ac.in}

\vspace{6mm}

${}^1${\small \sl Department of Physics and Astronomy} \\
{\small \sl Lehman College, City University of New York, Bronx, NY 10468, USA}

\vspace{2mm}

${}^2${\small \sl Department of Mathematics and} \\
{\small \sl Haifa Research Center for Theoretical Physics and Astrophysics} \\
{\small \sl University of Haifa, Haifa 31905, Israel}

\vspace{2mm}

${}^3${\small \sl Discipline of Physics} \\
{\small \sl Indian Institute of Technology Indore} \\
{\small \sl Khandwa Road 453552 Indore, India}

\end{center}

\vspace{12mm}

\noindent
We compute modular Hamiltonians for excited states obtained by perturbing the vacuum with a unitary operator.  We use operator methods and work to first order in the strength of the perturbation.  For the most part we divide space in half and focus on perturbations generated by integrating a local operator $J$ over a null plane.  Local operators with weight $n \geq 2$ under vacuum modular flow produce an additional endpoint contribution to the modular Hamiltonian.  Intuitively this is because operators with weight $n \geq 2$ can move degrees of freedom from a region to its complement.  The endpoint contribution is an integral of $J$ over a null plane.  We show this in detail for stress tensor perturbations in two dimensions, where the result can be verified by a conformal transformation, and for scalar perturbations in a CFT.  This lets us conjecture a general form for the endpoint contribution that applies to any field theory divided into half-spaces.

\end{titlepage}
\setcounter{footnote}{0}
\renewcommand\thefootnote{\mbox{\arabic{footnote}}}

\hrule
\tableofcontents
\bigskip
\hrule

\addtolength{\parskip}{8pt}
\section{Introduction}
For a CFT in its ground state the modular Hamiltonian for a spherical region is known \cite{Hislop:1981uh,Casini:2011kv}.
We're interested in the modular Hamiltonian for an excited state, obtained by acting on the vacuum
with a unitary operator.
\be
\label{state}
\vert \psi \rangle = \exp \Big[{-i \epsilon \int_\Sigma d^{d-1} x \, f({\bf x}) {\cal O}({\bf x})}\Big] \vert 0 \rangle
\ee
Here ${\cal O}({\bf x})$ is a generic Hermitian operator, $f({\bf x})$ is a real-valued function, and $\epsilon$ is
an expansion parameter.  At this stage the choice of hypersurface $\Sigma$ doesn't matter.  We'll start out thinking of $\Sigma$ as a spatial slice, but later
it will be convenient to choose $\Sigma$ to be a null plane.

Dividing space into a region $A$ and its complement $\bar{A}$ the reduced density matrices and subregion modular Hamiltonians are
\bea
\rho_A & = & e^{-H_A} = {\rm Tr}_{\bar{A}} \, \big(\vert \psi \rangle \langle \psi \vert\big) \\
\nonumber
\rho_{\bar{A}} & = & e^{-H_{\bar{A}}} = {\rm Tr}_A \, \big(\vert \psi \rangle \langle \psi \vert\big)
\eea
The full or extended modular Hamiltonian, which we denote $\Hext$, is the difference
\be
\Hext = H_A - H_{\bar{A}}
\ee
It's related to the modular operator by $\Delta = e^{-\Hext} = \rho_A \otimes \rho_{\bar{A}}^{-1}$.  Only the extended modular Hamiltonian is well-defined in the continuum.
Although subregion modular Hamiltonians make sense for systems with discrete degrees of freedom, they are not well-defined operators in the continuum.  This and related
matters are discussed in \cite{Witten:2018lha}.

How does the extended modular Hamiltonian $\Hext$ for the perturbed state differ from the extended modular Hamiltonian $\Hext^{(0)}$ for the vacuum?
Here's a naive and (as we'll see) incorrect argument.  Consider the unitary transformation appearing in (\ref{state}).
\be
\label{transformation}
U = e^{-i \epsilon G} = e^{-i \epsilon \int d^{d-1} x \, f({\bf x}) {\cal O}({\bf x})}
\ee
Naively the integral splits into an integral over $A$ and an integral over $\bar{A}$, which means the generator $G$ has the form
$G = G_A \otimes \identity_{\bar{A}} + \identity_A \otimes G_{\bar{A}}$.  Since $G_A$ and $G_{\bar{A}}$ commute, it would seem
the unitary transformation factors into
\be
\label{NaiveFactorization}
U = U_A \otimes U_{\bar{A}}
\ee
Given this form for $U$ it's easy to see that the reduced density matrices for the state $U \vert 0 \rangle$ are
$\rho_A = U_A \rho_A^{(0)} U_A^\dagger$ and $\rho_{\bar{A}} = U_{\bar{A}} \rho_{\bar{A}}^{(0)} U_{\bar{A}}^\dagger$,
which suggests that the perturbed modular operator and modular Hamiltonian should be
\be
\label{NaiveHmod}
\Delta = \rho_A \otimes \rho_{\bar{A}}^{-1} = U \Delta^{(0)} U^\dagger\,, \qquad \Hext = U \Hext^{(0)} U^\dagger\,.
\ee
The first-order change in the modular Hamiltonian would then be given by
\be
\label{deltaHcommutator}
\delta \Hext^{\rm commutator} = - i \epsilon [G,\Hext^{(0)}]
\ee

The problem with this argument is that generically the factorization (\ref{NaiveFactorization}) is not correct in field theory.  It would be valid in discrete spin or lattice models,
if the analog of $U$ is taken to be a tensor product of unitary transformations at each lattice site.  It would be valid in field theory if the function $f({\bf x})$ vanished in a neighborhood of
the ``endpoints,'' by which we mean the surface that separates $A$ from $\bar{A}$.  But generically $f$ won't vanish at the endpoints and there's no reason to trust the factorization
(\ref{NaiveFactorization}).

In what follows we'll show that a breakdown of (\ref{NaiveFactorization}) indeed invalidates the naive expression (\ref{NaiveHmod}) for $\Hext$.  To first order in $\epsilon$ we'll
identify a specific endpoint contribution $\delta \Hext^{\rm endpoint}$ which must be added to (\ref{deltaHcommutator}) to obtain the correct modular Hamiltonian.  Our explicit calculations are for perturbations
by the stress tensor in two dimensions and scalar primaries in general dimensions, but we're able to formulate the following conjecture for generic perturbations
that should hold in any field theory (not necessarily conformal):
\begin{center}
\fbox{\,\parbox{16cm}{\small Divide space into $\lbrace x > 0 \rbrace \cup \lbrace x < 0 \rbrace$ and perturb the vacuum by acting with
\be
G = \int_{-\infty}^\infty dx^+ d^{d-2}x_\perp f(x^+,{\bf x}_\perp) J^{(n)}(x^+,0,{\bf x}_\perp)
\ee
Here $x^\pm = t \pm x$ and $J^{(n)}$ is a local operator that transforms with weight $n$ under a Lorentz boost (for example $T_{++}$ has weight 2).
Note that the perturbation acts on the null plane $x^- = 0$.  Then the endpoint contribution which must be added to (\ref{deltaHcommutator}) to obtain the
change in the extended modular Hamiltonian is
\be
\label{deltaHext}
\delta \Hext^{\rm endpoint} = \left\lbrace
\begin{array}{ll}
- 2 \pi \epsilon (n-1) \int d^{d-2} x_\perp \, f(0,{\bf x}_\perp) \int_{-\infty}^\infty dx^+ J^{(n)}(x^+,0,{\bf x}_\perp) \quad & \hbox{\rm for $n = 2,3,4,\ldots$} \\[5pt]
\quad 0 & \hbox{\rm for $n = \ldots,-1,0,1$}
\end{array}\right.
\ee}\,}
\end{center}
Curiously endpoint contributions only arise for weight $n \geq 2$.  To provide intuition for this, we will argue that only operators of
weight 2 or greater can move local degrees of freedom from $A$ to $\bar{A}$ or visa versa along the $x^+$ Rindler horizon.

Related works have appeared in the literature, including studies of entanglement and bulk reconstruction for excited states \cite{Rosenhaus:2014woa,Rosenhaus:2014zza,Speranza:2016jwt,Belin:2018juv,Belin:2019mlt} and the behavior of modular Hamiltonians and entanglement in excited states and under shape deformations \cite{Allais:2014ata,Lashkari:2015dia,Faulkner:2015csl,Sarosi:2016oks,Faulkner:2016mzt,Casini:2017roe,Sarosi:2017rsq,Lewkowycz:2018sgn,Lashkari:2018oke,deBoer:2019uem,Rosso:2019txh,Balakrishnan:2020lbp,Arias:2020qpg,Rosso:2020cub}.  In contrast to these previous works, we consider excited states produced by perturbing the vacuum
by a generic operator integrated over a null plane.  However shape deformations, in particular, generate an endpoint contribution built from the stress tensor identical to what we
obtain here \cite{Faulkner:2016mzt}.  Null-integrated operators have also been considered in \cite{Kravchuk:2018htv,Kologlu:2019mfz}.

As an outline, in section \ref{sect:conformal} we consider perturbations by the stress tensor in $1+1$ dimensions.  In this case, as shown by Das and Ezhuthachan \cite{Das:2018ojl},
the exact modular Hamiltonian can be obtained using a conformal transformation.  We expand their results to first order in $\epsilon$ and identify the endpoint contributions.
This shows that they're non-vanishing and gives us a target to aim for.  In section \ref{sect:perturbation} we collect results in the literature to obtain an expression for the first-order correction to the
modular Hamiltonian for a rather general class of perturbations.  In section \ref{sect:endpoint} we show that the general formalism of section \ref{sect:perturbation}
can be applied to perturbations by the stress tensor in two dimensions and, with an appropriate treatment of endpoints, reproduces the results obtained from a conformal transformation in section \ref{sect:conformal}.
In section \ref{sect:scalar} we present an analogous calculation of the (vanishing) endpoint contribution for scalar perturbations, and in section \ref{sect:general} we present a general conjecture for the endpoint
contribution to the subregion modular Hamiltonian.

Supporting information is collected in the appendices: the construction of modular Hamiltonians using conformal transformations is reviewed in appendix \ref{appendix:excited}, results for the perturbation
series are collected in appendix \ref{appendix:perturbation}, the steps outlined in section \ref{sect:general} are illustrated in a 2-D CFT in appendix \ref{appendix:generaln},
properties of OPEs and commutators are considered in appendix \ref{appendix:OPE}, and CFT normalizations are discussed in appendix \ref{appendix:conventions}.

Our conventions are as follows.  Light-front coordinates are denoted $x^\pm = t \pm x$.  In two dimensions, to agree with the CFT conventions of \cite{Ginsparg:1988ui}, instead of the usual stress tensor we sometimes work with
\be
T(x^+) = - 2 \pi T_{++}(x^+)
\ee
Finally for a division into half-spaces the modular Hamiltonian $\Hext$ is related to a Lorentzian boost or Euclidean angular rotation generator $K$ by $\Hext = 2 \pi K$.

\section{Conformal transformations and endpoint contributions\label{sect:conformal}}
In this section we consider a two-dimensional Lorentzian CFT, perturbed by applying the stress tensor to the vacuum.  The modular Hamiltonian can be obtained exactly with the help of a conformal transformation.
This is a concrete setting where we can independently argue for and identify the endpoint contributions.  This will provide a valuable check on our later work, where we reproduce these
results by more general methods.

For a CFT in its vacuum state the extended modular Hamiltonian for an interval $(u,v)$ is known \cite{Hislop:1981uh,Casini:2011kv}.
\be
\label{Hmod0}
\Hext^{(0)}_{(u,v)} = 2 \pi \int_{-\infty}^\infty dz \, {(v - z)(z - u) \over v - u} \, T_{++}(z) + (\hbox{\rm right-movers})
\ee
We'll suppress the right-moving contribution in what follows.  For an excited state, obtained from the vacuum by applying a conformal transformation, the extended
modular Hamiltonian for the same interval is \cite{Das:2018ojl}
\be
\label{Hmodg}
\Hext_{(u,v)} = 2\pi \int_{-\infty}^\infty dz \, {(g(v) - g(z)) (g(z) - g(u)) \over g'(z)(g(v) - g(u))} \, T_{++}(z)
\ee
Here $g$ is a function that parametrizes the state.
The derivation of this result is reviewed in appendix \ref{appendix:excited}.

We're interested in a linearized perturbation by the stress tensor, or in other words an infinitesimal conformal transformation, with generator
\be
\label{G}
G = \int dz \, f(z) T_{++}(z)
\ee
So we set $g(z) = z + \epsilon f(z)$ and expand (\ref{Hmodg}) to first order in $\epsilon$ to find
\bea
\nonumber
\delta \Hext_{(u,v)} & = & \Hext_{(u,v)} - \Hext^{(0)}_{(u,v)} \\
\nonumber
& = & 2\pi \int_{-\infty}^\infty dz \, T_{++}(z) \Big[- \epsilon f'(z) {(v - z)(z - u) \over v - u} + \epsilon f(z) {u + v - 2 z \over v - u} \\
\label{ConformalDeltaH}
&& \qquad + \epsilon f(v) \left({z - u \over v - u}\right)^2 - \epsilon f(u) \left({v - z \over v - u}\right)^2\Big]
\eea
The second line is particularly interesting.  It vanishes if $f(u) = f(v) = 0$, that is, if the conformal transformation leaves the endpoints of the interval invariant.
In what follows we will argue that
\begin{itemize}
\item
The first line of (\ref{ConformalDeltaH}) can be identified as a commutator contribution.  That is, it's what one would obtain by expanding the naive result
(\ref{NaiveHmod}) to first order in $\epsilon$.
\item
The second line of (\ref{ConformalDeltaH}) can be identified as an endpoint contribution.  It represents an additional term, not present in (\ref{NaiveHmod}),
which arises from a careful treatment of the endpoints.
\end{itemize}

A first check of these claims is to note that the naive result (\ref{NaiveHmod}) suggests that the change in the modular Hamiltonian should be
$- i \epsilon [G, \Hext^{(0)}_{(u,v)}]$.  It's straightforward to evaluate this commutator using the Virasoro algebra
\be
\label{Virasoro}
i[T_{++}(x),T_{++}(z)] = {c \over 24 \pi} \delta^{\prime\prime\prime}(x-z) - 2 T_{++}(z) \delta^\prime(x-z) + \partial_z T_{++}(z) \delta(x-z)
\ee
Ignoring the central term one finds that this reproduces the first line of (\ref{ConformalDeltaH}) but misses the endpoint contribution.

As further evidence, and for future reference, it's useful to zoom in on one of the endpoints.  To do this it suffices to consider a division into half-spaces
$\lbrace x > 0 \rbrace \cup \lbrace x < 0 \rbrace$, so we set $u = 0$ and $v = \infty$ in (\ref{ConformalDeltaH}) to obtain
\be
\label{ConformalDeltaH_half_space}
\delta \Hext_{(0,\infty)} = 2 \pi \int_{-\infty}^\infty dz \, T_{++}(z) \big[-\epsilon z f'(z) + \epsilon f(z) - \epsilon f(0) \big]
\ee
We are proposing that the last term is the endpoint contribution.  To get a better understanding of this consider a transformation which is supported on an interval $(-b,a)$ and takes a constant
value $f_0$ in that interval.
\be
\label{f}
f(x) = f_0 \theta(x+b) - f_0 \theta(x-a)
\ee
\centerline{\includegraphics[width=5cm]{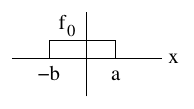}}
Here $a$ and $b$ are positive quantities.  We'll frequently consider sending $a,b \rightarrow 0^+$ but in all calculations we will keep them finite.  For such a transformation
the perturbation to the modular Hamiltonian is
\be
\label{deltaHf0}
\delta \Hext_{(0,\infty)} = 2 \pi \epsilon f_0 \int_{-\infty}^\infty dz \, T_{++}(z) \big[b \delta(z+b) + a \delta(z-a) + \theta(z+b) - \theta(z-a) - 1 \big]
\ee
Note that sending $a,b \rightarrow 0^+$ gives
\be
\label{deltaHextlimit}
\delta \Hext_{(0,\infty)} \rightarrow - 2 \pi \epsilon f_0 \int_{-\infty}^\infty dz \, T_{++}(z)
\ee
In this limit $\delta \Hext_{(0,\infty)}$ remains non-trivial and is given purely by the endpoint contribution.  This is a bit remarkable because from (\ref{f}) and (\ref{G}) it naively appears that
in this limit we should have $f(x) \rightarrow 0$ and $G \rightarrow 0$.  The fact that a non-trivial endpoint contribution survives shows that the modular Hamiltonian
is sensitive to the details of what happens near the endpoint.

\section{Modular Hamiltonians for perturbed states\label{sect:perturbation}}
In this section we give an expression for the first-order change in the subregion modular Hamiltonian for a rather generic perturbation to the state.
The treatment, which builds on results in the literature, is somewhat formal and ignores endpoint subtleties.  We summarize the results here
and give further details in appendix \ref{appendix:perturbation}.

Given a state $\vert \psi \rangle = e^{-i \epsilon G} \vert 0 \rangle$, to first order in $\epsilon$ the change in the state is $\delta \vert \psi \rangle = - i \epsilon G \vert 0 \rangle$.
We assume the generator can be factored, $G = G_A \otimes G_{\bar{A}}$.\footnote{More generally we could allow $G$ to be a sum of such tensor products, and we will need
this generalization in section \ref{sect:endpoint}.  But to first order in perturbation theory one just sums the different contributions to $\delta H_A$.}  Then the first-order change in the modular Hamiltonian for subregion $A$ is
\be
\label{deltaHA}
\delta H_A = {i \epsilon \over 2} \int_{-\infty}^\infty {ds \over 1 + \cosh s} \left(G_A\big\vert_{s - i \pi} \widetilde{G}_{\bar{A}} \big\vert_s
- \widetilde{G}_{\bar{A}}\big\vert_s G_A\big\vert_{s+i\pi}\right)
\ee
There's an analogous expression for $\delta H_{\bar{A}}$ that can be obtained by interchanging $A \leftrightarrow \bar{A}$.
Here $\widetilde{G}_{\bar{A}}$ denotes the mirror operator to $G_{\bar{A}}$.  As discussed in appendix \ref{appendix:perturbation} it's given by modular conjugation, $\widetilde{G}_{\bar{A}} = J G_{\bar{A}} J$, which for a division into half-spaces is the same as conjugation by a CPT transformation.
The integral is over vacuum modular flow, which we denote by
\be
{\cal O}\big\vert_s = \Delta^{-{i s \over 2 \pi}} {\cal O} \Delta^{i s \over 2 \pi}
\ee
where $\Delta$ is the vacuum modular Hamiltonian.  For a half-space $\Delta = e^{-2 \pi K}$ is the exponential of a Lorentz generator \cite{Witten:2018lha,Bisognano:1976za}
which means we can think of modular flow as a Lorentz boost.
\be
{\cal O}\big\vert_s = e^{i K s} {\cal O} e^{-i K s}
\ee
To use (\ref{deltaHA}) we'll need to make sense of a complex Lorentz boost.  We'll tackle this in the next section.

\section{Stress tensor perturbations in two dimensions\label{sect:endpoint}}
In this section we consider perturbations by the stress tensor in a two-dimensional CFT.  Our goal is to understand (\ref{ConformalDeltaH}), including endpoint contributions, from the perspective of the expansion developed in section \ref{sect:perturbation}.  Equation (\ref{deltaHA}) gives us an expression for the change in the subregion modular Hamiltonian $\delta H_A$, so that's the quantity we'll focus on.  Similar results hold for $\delta H_{\bar{A}}$.

Rather than work in full generality, for simplicity we consider a division of space into $A = \lbrace x > 0 \rbrace$ and $\bar{A} = \lbrace x < 0 \rbrace$.  We break the generator $G$ into three pieces.
\be
\label{Gpieces}
G = \int_{-\infty}^{-b} dx f(x) T_{++}(x) + \int_{-b}^a dx f(x) T_{++}(x) + \int_a^\infty dx f(x) T_{++}(x)
\ee
Here $a$ and $b$ are small but fixed positive quantities.  Since we're working to first order in $\epsilon$ we can consider the different pieces of $G$ separately and add their contributions.

\subsection{First term}
The first term in (\ref{Gpieces}) can be decomposed as $G_A \otimes G_{\bar{A}}$ with
\be
G_A = \identity_A \qquad\quad G_{\bar{A}} = \int_{-\infty}^{-b} dx f(x) T_{++}(x)
\ee
Plugging this into (\ref{deltaHA}) rather trivially gives $\delta H_A = 0$.  This is no surprise: this term only acts on $\bar{A}$ so it leaves the reduced density matrix $\rho_A$ invariant.

\subsection{Third term}
The third term in (\ref{Gpieces}) has the decomposition
\be
\label{term3generator}
G_A = \int_a^\infty dx f(x) T_{++}(x) \qquad\quad G_{\bar{A}} = \identity_{\bar{A}}
\ee
Substituting this in (\ref{deltaHA}) gives
\be
\delta H_A = {i \epsilon \over 2} \int_{-\infty}^\infty {ds \over 1 + \cosh s} \left(G_A\big\vert_{s - i \pi} - G_A\big\vert_{s+i\pi}\right)
\ee
In the second term make the change of variables $s = \tilde{s} - 2 \pi i$.  Then the integrands in the two terms are the same and we have
\be
\delta H_A = {i \epsilon \over 2} \int_a^\infty dx f(x) \, \left[\int_{-\infty}^\infty + \int_{\infty + 2\pi i}^{-\infty + 2\pi i}\right] {ds \over 1 + \cosh s} \, \Delta^{-{1 \over 2} - {i s \over 2 \pi}} \, T_{++}(x) \, \Delta^{{1 \over 2} + {i s \over 2 \pi}}
\ee
We'll assume we can close the contour near ${\rm Re} \, s = \pm \infty$ and that the only contribution to the integral comes from the pole at $s = i \pi$ in ${1 \over 1 + \cosh s}$.  That is,
we assume we can evaluate the integral using
\be
\oint {ds \over 1 + \cosh s} \, g(s) = - 4 \pi i g'(s = i \pi)
\ee
In our case the derivative gives a commutator with the unperturbed modular Hamiltonian, and the modular operators disappear when $s = i \pi$.  So we're left with the simple result that
\be
\label{term3deltaHmod}
\delta H_A = - i \epsilon \int_a^\infty dx \, f(x) \, [ T_{++}(x), \Hext^{(0)} ]
\ee
Again this is no surprise: the naive expression (\ref{NaiveHmod}) implies that $H_A = U_A H^{(0)}_A U_A^\dagger$, which at first order in $\epsilon$ gives the commutator (\ref{term3deltaHmod}).  To evaluate the commutator we use
\be
\Hext^{(0)}_{(0,\infty)} = 2 \pi \int_{-\infty}^\infty dz \, z T_{++}(z)
\ee
and the Virasoro algebra (\ref{Virasoro}).  Ignoring the central term we find
\be
\label{PerturbativeDeltaHA}
\delta H_A = - 2 \pi \epsilon a f(a) T_{++}(a) + 2 \pi \epsilon \int_a^\infty dz \, T_{++}(z) \big(-z f'(z) + f(z)\big)
\ee

We'd like to show that this agrees with the result (\ref{ConformalDeltaH_half_space}) which we obtained using a conformal transformation.
To do this we substitute $f(z) \rightarrow f(z) \theta(z-a)$ in (\ref{ConformalDeltaH_half_space}), where the step function corresponds to keeping just the third term in (\ref{Gpieces}).  Then the conformal result (\ref{ConformalDeltaH_half_space}) implies
\bea
&& \delta H_A = - 2 \pi \epsilon a f(a) T_{++}(a) + 2 \pi \epsilon \int_a^\infty dz \, T_{++}(z) \big(-z f'(z) + f(z)\big) \\
\nonumber
&& \delta H_{\bar{A}} = 0
\eea
in agreement with (\ref{PerturbativeDeltaHA}).  Note that since $a > 0$ there is no endpoint contribution to this expression.  That is, only the first line (the commutator terms) in (\ref{ConformalDeltaH}) contribute.  Also note that the delta function contribution
$- 2 \pi \epsilon a f(a) T_{++}(a)$ to $\delta H_A$ will cancel against a similar term that appears in (\ref{HAcommutator}) when the different parts of (\ref{Gpieces}) are assembled.

\subsection{Middle term}\label{subsec:2dT}
Finally we consider the middle term in (\ref{Gpieces}).  Here things get more subtle.  Although $a$ and $b$ are finite, let's imagine they're small enough that we can treat
$f(x)$ as approximately constant over the interval $(-b,a)$, with $f(x) \approx f_0$.  So we work with
\bea
\nonumber
G & = & f_0 \int_{-b}^a dx \, T_{++}(x) \\
\label{T++pert}
& = & - {f_0 \over 2 \pi} \int_{-b}^a dx \, T(x)
\eea
where we switched to the CFT normalization discussed in appendix \ref{appendix:conventions}, $T = - 2 \pi T_{++}$.

It's not obvious that $G$ can be decomposed into operators on $A$ and $\bar{A}$.  To gain some insight we consider the vacuum correlator $\langle G T(y) \rangle$.\footnote{The correlators
we'll work with below, such as (\ref{GT}) and (\ref{TTT}), should be understood as Wightman functions.  Strictly speaking
they should be defined with appropriate $i \epsilon$ prescriptions.  But in what follows we won't have to deal with lightcone singularities, so the $i \epsilon$ prescriptions won't matter and
we can ignore this subtlety.\label{footnote:Wightman}}
Given the stress tensor two-point function $\langle T(x) T(y) \rangle = {c / 2 \over (x - y)^4}$ it's straightforward to compute\footnote{As in footnote \ref{footnote:Wightman} this is a Wightman function.  Strictly speaking it's defined by giving $a$ and $-b$ small negative imaginary parts.
But for our purposes it's sufficient to note that the integral is well-defined when $y \not\in [-b,a]$, and there's no obstacle to continuing the result into the region $-b < y < a$.}
\be
\label{GT}
\langle G T(y) \rangle = - {f_0 c \sigma \over 6 \pi} \, {3(y - \delta)^2 + \sigma^2 \over (y-a)^3 (y+b)^3}
\ee
where
\be
\sigma = {a + b \over 2} \qquad\quad \delta = {a - b \over 2}
\ee
Note that (\ref{GT}) has singularities at $y = a$ and $y = -b$.  This suggests that, although $G$ itself isn't localized to either $A$ or $\bar{A}$, it can be decomposed into tensor products
of local operators at $a$ with local operators at $-b$.

To identify these local operators consider the 3-point function
\be
\label{TTT}
\langle T(a) T(-b) T(y) \rangle = {c \over (a+b)^2 (a-y)^2 (-b-y)^2}
\ee
There's a useful expansion
\be
\label{expansion}
\sum_{n = 0,2,4,\ldots} \alpha_n \, \sigma^n (\partial_a - \partial_b)^n {1 \over (a-y)^2 (-b-y)^2} = {3(y-\delta)^2 + \sigma^2 \over 3 (y-a)^3 (y+b)^3}
\ee
The coefficients $\alpha_n$ can be determined by organizing both sides as an expansion in powers of ${\sigma^2 \over (y-\delta)^2}$ and demanding agreement
order-by-order.  The first few coefficients are
\be
\alpha_0 = 1,\quad \alpha_2 = {1 \over 15},\quad \alpha_4 = - {1 \over 525},\quad \alpha_6 = {2 \over 23625},\quad \ldots
\ee
The details of these coefficients won't matter, as long as one accepts that there's an expansion of the form (\ref{expansion}).
It means that we can write
\be
\label{GTexpansion}
\langle G T(y) \rangle = -{2 f_0 \sigma^3 \over \pi} \sum_n \alpha_n \sigma^n (\partial_a - \partial_b)^n \langle T(a) T(-b) T(y) \rangle
\ee
So at least inside a correlator with another stress tensor, $G$ can be decomposed into an infinite sum of tensor products of local operators at $a$ with local operators at $-b$.  The local operators involve
arbitrary derivatives of the stress tensor.

This result can be lifted to an operator identity\footnote{The central term is ${f_0 c \over 16 \pi \sigma} \identity$, as can be seen by requiring $\langle G \rangle = 0$.}
\be
\label{T++operatorid}
G = -{2 f_0 \sigma^3 \over \pi} \sum_n \alpha_n \sigma^n (\partial_a - \partial_b)^n T(a) T(-b) + (\hbox{\rm central term})
\ee
which holds inside any correlator.  To see this take the correlator of (\ref{T++operatorid}) with an operator $X$.  If $X$ is not in the conformal family of the identity then the conformal Ward identities
\cite{Belavin:1984vu} imply that both sides vanish.  If $X$ is in the conformal family of the identity then both sides are equal as a consequence of (\ref{GTexpansion}); we chose the central term so the two sides agree even
for $X = \identity$.  Since the two sides agree for any $X$, the operator identity (\ref{T++operatorid}) should hold inside any correlator.

Please note that it's important to keep the full sum in (\ref{GTexpansion}).  From (\ref{GT}) it's tempting to think that as $a,b \rightarrow 0^+$
we should have
\be
\langle G T(y) \rangle \rightarrow - {f_0 c \sigma \over 2 \pi y^4}
\ee
This behavior could be reproduced by keeping just the $n = 0$ term in (\ref{GTexpansion}) and taking the OPE limit $a,b \rightarrow 0^+$.  But when computing $\delta H_A$ we need to trace over the half-space
$x < 0$.  When doing this we need to hold $a$ and $b$ fixed and finite, and we need to accurately capture the effects of $G$ on the state even in the interval between $-b$ and $a$.  So it's important to retain the
full sum in (\ref{GTexpansion}), since this allows us to reproduce (\ref{GT}) exactly even for $-b < y < a$.\footnote{A concrete manifestation of the role of the sum is that keeping just the $n = 0$ term in
(\ref{HATintegral}) gives an expression which -- unlike (\ref{HAT}) -- is ill-defined as $a,b \rightarrow 0^+$.}  By keeping the full sum we're able to compute the change
in the modular Hamiltonian without introducing a regulator and without modifying the state in any way.  This is important to avoid mistakes because,
as pointed out below (\ref{deltaHextlimit}), the modular Hamiltonian is quite sensitive to the details of the state near the endpoint.

Given the expansion of $G$ into a sum of operators on $A$ times operators on $\bar{A}$, the next step is to use the representation of $\delta H_A$ given in (\ref{deltaHA}).
This gives
\bea
\delta H_A & = & - {i \epsilon f_0 \sigma^3 \over \pi} \int_{-\infty}^\infty {ds \over 1 + \cosh s} \sum_n \alpha_n \, \sigma^n \sum_{k=0}^n \left({n \atop k}\right) \bigg[ \\
\nonumber
& & \partial_+^{n-k} T(a) \big\vert_{s-i\pi} \, \widetilde{\partial_+^k T}(-b) \big\vert_s - \widetilde{\partial_+^k T}(-b) \big\vert_s \, \partial_+^{n-k} T(a) \big\vert_{s+i\pi} \bigg]
\eea
To construct the mirror operators we use CPT conjugation.  CPT acts on light-front coordinates by $x^\pm \rightarrow - x^\pm$, so the mirror operators are given by
\be\label{eq:2dcpt1}
\widetilde{\partial_+^k T_{++}}(-b) = (-1)^k \partial_+^k T_{++}(b)
\ee
or equivalently
\be\label{eq:2dcpt2}
\widetilde{\partial_+^k T}(-b) = (-1)^k \partial_+^k T(b)
\ee
This leads to
\be
\label{deltaHApert}
\delta H_A = - {i \epsilon f_0 \sigma^3 \over \pi} \int_{-\infty}^\infty {ds \over 1 + \cosh s} \sum_n \alpha_n \, \sigma^n (\partial_a - \partial_b)^n \Big[
T(a) \big\vert_{s-i\pi} T(b) \big\vert_s - T(b) \big\vert_s T(a) \big\vert_{s+i\pi} \Big]
\ee
Next we need to make sense of complex modular flow.  Given the pitfalls discussed in section 4.2 of \cite{Witten:2018lha}, rather than try this at the operator level we insert $\delta H_A$ in a correlator with another stress tensor.  Thus, inserting the additional stress tensor at $y < 0$ to avoid singularities, we consider
\bea
\langle \delta H_A T(y) \rangle & = & - {i \epsilon f_0 \sigma^3 \over \pi} \int_{-\infty}^\infty {ds \over 1 + \cosh s} \sum_n \alpha_n \, \sigma^n (\partial_a - \partial_b)^n \Big[ \\
\nonumber
& & \langle T(a) \big\vert_{s-ir} T(b) \big\vert_s T(y) \rangle - \langle T(b) \big\vert_s T(a) \big\vert_{s+ir} T(y) \rangle\Big]
\eea
Our approach will be to start at $r = 0$, where the correlator is well-defined, and analytically continue to $r = \pi$.  Since modular flow is a Lorentz boost it acts on the stress tensor by\footnote{In general for
a Lorentz boost $e^{i K s} T_{\alpha\beta}(x) e^{-iKs} = {\partial x'^\mu \over \partial x^\alpha} {\partial x'^\nu \over \partial x^\beta} T_{\mu\nu}(x')$.  On light-front coordinates the boost acts by
$x'^+ = e^s x^+$, $x'^- = e^{-s} x^-$ which leads to (\ref{T++flow}).}
\be
\label{T++flow}
T_{++}(x^+) \big\vert_s = e^{2 s} T_{++}(e^s x^+)
\ee
or equivalently
\be
\label{Tflow}
T(x^+) \big\vert_s = e^{2 s} T(e^s x^+)
\ee
We'll assume this makes sense even for complex $s$.  The stress tensor 3-point function is given in (\ref{TTT}).  Using (\ref{Tflow}) and letting $z = e^s$ we obtain
\bea
\nonumber
\langle \delta H_A T(y) \rangle & = & -{2 i \epsilon f_0 \sigma^3 c \over \pi} \sum_n \alpha_n \, \sigma^n (\partial_a - \partial_b)^n \bigg[{1 \over (a - e^{ir} b)^2} \int_0^\infty {dz \over (z+1)^2} \, {z^2 \over (e^{-ir} az - y)^2 (bz - y)^2} \\
& & - {1 \over (a - e^{-ir} b)^2} \int_0^\infty {dz \over (z+1)^2} \, {z^2 \over (e^{ir}az - y)^2(bz-y)^2}\bigg]
\eea
The first integral has poles at $z = -1,\, y/b,\, e^{ir} y / a$.  Remembering that $y < 0$, as $r$ increases from $0$ to $\pi$ the last pole moves counterclockwise and hits the contour of integration
from below.  The second integral has poles at $z = -1,\, y/b,\, e^{-ir} y / a$.  As $r$ increases the last pole moves clockwise and hits the contour of integration from above.

\centerline{\includegraphics[width=5cm]{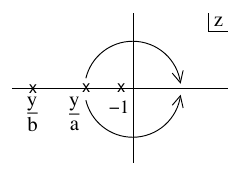}}

\noindent
Thus when $r = \pi$
\bea
\nonumber
\langle \delta H_A T(y) \rangle & = & -{2 i \epsilon f_0 \sigma^3 c \over \pi} \sum_n \alpha_n \, \sigma^n (\partial_a - \partial_b)^n \Big[{1 \over (a + b)^2} \landupint_0^\infty dz {z^2 \over (z+1)^2 (az + y)^2 (bz - y)^2} \\
\label{HATintegral}
& & - {1 \over (a + b)^2} \landdownint_0^\infty dz {z^2 \over (z+1)^2(az + y)^2(bz-y)^2}\Big]
\eea
where the integrals pass above or below the pole at $z = -y/a$ as indicated on the integral symbols.  Fortunately at this point we can use the identity (\ref{expansion}) to do the sum (we never needed the coefficients $\alpha_n$!).  After doing the sum
the difference between the two integrals picks up the residue of the pole at $-y/a$ and gives
\be
\label{HAT}
\langle \delta H_A T(y) \rangle = - {\epsilon f_0 c \over 6} \, {y + 2 a \over (y - a)^4}
\ee

Rather remarkably this matches what we'd expect based on the results of section \ref{sect:conformal}.  From the $x > 0$ piece of (\ref{deltaHf0}) we'd expect
\be
\label{deltaHAconf}
\delta H_A = - \epsilon f_0 a T(a) + \epsilon f_0 \int_a^\infty dx \, T(x)
\ee
which as in (\ref{ConformalDeltaH}) can be decomposed into commutator and endpoint contributions (except that now the endpoint is at $x = 0$).
\bea
\label{HAcommutator}
&& \delta H_A^{\rm commutator} = - \epsilon f_0 a T(a) - \epsilon f_0 \int_0^a dx \, T(x) \\
\label{HAendpoint}
&& \delta H_A^{\rm endpoint} = \epsilon f_0 \int_0^\infty dx \, T(x)
\eea
To show that these are indeed the operators appearing in (\ref{HAT}) we use the 2-point function $\langle T(x) T(y) \rangle = {c / 2 \over (x - y)^4}$ to calculate (remembering that $y < 0$)
\bea
\nonumber
&& \langle \delta H_A^{\rm commutator} T(y) \rangle = {\epsilon f_0 c \over 6} \left({1 \over y^3} - {y + 2 a \over (a - y)^4}\right) \\
&& \langle \delta H_A^{\rm endpoint} T(y) \rangle = - {\epsilon f_0 c \over 6 y^3}
\eea
The sum of these two correlators agrees with (\ref{HAT}).  Note that as $a \rightarrow 0$ only the endpoint contribution survives.

The computations above evaluate the endpoint contribution for a Rindler half-space. But in principle, our methods can be implemented for any finite subregion $(u,v)$. The main differences are that the modular-conjugated operator $\widetilde{G}_{\bar{A}}$ is no longer given by
CPT conjugation (as in \eqref{eq:2dcpt1}) and vacuum modular flow is no longer given by a Lorentz boost (as in (\ref{T++flow})).  Alternatively for a CFT we can obtain the
finite-subregion result using a global conformal transformation
\begin{equation}
\label{GlobalConformalMap}
x^+=(v-u)\,\frac{\tilde{x}^+-u}{v-\tilde{x}^+}
\end{equation}
The half-space $0 < x^+ < \infty$ is the image of a finite region $(u,v)$ in $\tilde{x}^+$ variables.  Pulling back (\ref{Gpieces}) under this map as in
appendix \ref{appendix:excited}, we find an excited state of the same form but with $f(x^+)$ replaced by
\be
\tilde{f}(\tilde{x}^+) = \left({v - \tilde{x}^+ \over v - u\,\,}\right)^2 f(x^+)
\ee
Likewise pulling back the endpoint contribution \eqref{HAendpoint} and subtracting a similar contribution from $\bar{A}$ recovers the
$u$ endpoint contribution given in the second line of \eqref{ConformalDeltaH}, with the replacement $f(u) \rightarrow \tilde{f}(u)$.

\section{Scalar perturbations\label{sect:scalar}}
In this section we consider perturbing the state by a scalar primary operator ${\cal O}$.  We keep both the dimension of the operator $\Delta$ and the dimension of the CFT $d$ general.  Our goal is to determine
the endpoint contribution to the modular Hamiltonian.

The first step is to decide what form the perturbation should take.  In (\ref{T++pert}) we perturbed by the the stress tensor $T_{++}(x)$ on a spatial interval $-b < x < a$.  But $T_{++}$ only depends on the $x^+$ light-front
coordinate, so we could equally well think of the perturbation as acting on the null interval $-b < x^+ < a$.  When treating scalar perturbations, this observation motivates us to consider perturbations that act on a
null plane.  So we consider scalar perturbations generated by\footnote{We're indicating light-front coordinates by ${\cal O}(x^+,x^-,{\bf x}_\perp)$ with $ds^2 = - dx^+ dx^- + \vert d{\bf x}_\perp \vert^2$.}
\be
G = \int_{-b}^a dx^+ \, {\cal O}(x^+,0,0)
\ee
This choice of perturbation is not strictly necessary but simplifies many of the formulas that follow.

We want to decompose $G$ into operators on $A = \lbrace x > 0 \rbrace$ and $\bar{A} = \lbrace x < 0 \rbrace$.  Rather than work with $G$ directly we study the correlator $\langle G \, {\cal O}(y) \rangle$.
Using the two-point function $\langle {\cal O}(x) {\cal O}(y) \rangle = {1 \over \left((x-y)^2\right)^\Delta}$ it's straightforward to obtain
\be
\label{scalarGO}
\langle G \, {\cal O}(y) \rangle = - {1 \over (\Delta - 1) y^-} \left[{1 \over (a y^- + y^2)^{\Delta - 1}} - {1 \over (-b y^- + y^2)^{\Delta - 1}}\right]
\ee
This is singular whenever $y = (y^+,y^-,{\bf y}_\perp)$ is null separated from $(a,0,0)$ and $(-b,0,0)$,\footnote{Note that $\left((x^+,0,0) - y\right)^2 = x^+ y^- + y^2$.} which suggests that $G$ can be decomposed into a
sum of products of local operators at these points.  To make this precise we introduce a regulator to make these points spacelike separated.  So we introduce a parameter $\theta
\rightarrow 0^+$ and replace
\be
\label{regulator}
(a,0,0) \rightarrow (a,-\theta a,0) \qquad (-b,0,0) \rightarrow (-b,-\theta(-b),0)
\ee

\centerline{\includegraphics[width=5cm]{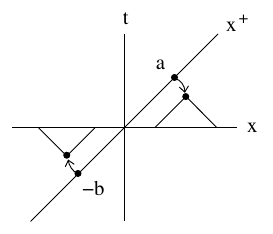}}

\noindent
This regulator makes the operator product non-singular.  Another advantage of this regulator is that the operator at $a$ unambiguously acts on $A$ while the operator at $-b$ unambiguously acts on $\bar{A}$.

Next we need to decide what operator product to use.  We will show that $G$ can be built from the symmetrized combination ${\cal O}(a) T_{++}(-b) + T_{++}(a) {\cal O}(-b)$.  To establish this
we consider the correlator
\be
C(a,b,y) = \langle \big[{\cal O}(a) T_{++}(-b) + T_{++}(a) {\cal O}(-b)\big] {\cal O}(y) \rangle
\ee
where
\be
{\cal O}(a) \equiv {\cal O}(a, -\theta a,0) \qquad T_{++}(a) \equiv T_{++}(a,-\theta a,0)
\ee
This can be evaluated using the three-point function\footnote{See for example (24) in \cite{Penedones:2016voo}.}
\bea
\nonumber
&& \langle {\cal O}(x_1) {\cal O}(x_2) T_{\mu\nu}(x_3) \rangle = C_{{\cal OO}T} {H_{\mu\nu} \over \left(x_{12}^2\right)^{\Delta - {d \over 2} + 1} \left(x_{13}^2\right)^{{d \over 2} - 1} \left(x_{23}^2\right)^{{d \over 2} - 1}} \\[5pt]
&& H_{\mu\nu} = V_\mu V_\nu - {1 \over d} g_{\mu\nu} V^2,\quad V^\mu = {x_{13}^\mu \over x_{13}^2} - {x_{23}^\mu \over x_{23}^2}
\eea
Retaining just the leading behavior as $\theta \rightarrow 0^+$, and denoting as before
\be
\sigma = {a + b \over 2} \qquad \delta = {a - b \over 2}
\ee
we find that there's an expansion\footnote{In this formula $(2 \sigma)^{d+1}$ really means $2 \sigma ((2\sigma)^2)^{d/2}$, which changes sign under $\sigma \rightarrow - \sigma$.\label{footnote:odd}}
\be
\label{GOexpansion}
\langle G {\cal O}(y) \rangle = {2 \theta^{{d \over 2} - 1} (2\sigma)^{d + 1} \over C_{{\cal OO}T}} \sum_{n = 0,2,4,\ldots} \beta_n \sigma^n (\partial_a - \partial_b)^n C(a,b,y)
\ee
The first few coefficients are
\be
\beta_0 = 1, \qquad \beta_2 = - {2 \Delta^2 - (6 d +10) \Delta + 3 (d+2)^2 \over 6 \Delta (\Delta + 1)}, \qquad \cdots
\ee
This identity, reminiscent of (\ref{GTexpansion}), can be established by expanding both sides of (\ref{GOexpansion}) in powers of ${\sigma y^- \over \delta y^- + y^2}$.

We want to lift this to an operator identity
\be
\label{scalaroperatorid}
G = {2 \theta^{{d \over 2} - 1} (2\sigma)^{d + 1} \over C_{{\cal OO}T}} \sum_{n = 0,2,4,\ldots} \beta_n \sigma^n (\partial_a - \partial_b)^n \big({\cal O}(a) T_{++}(-b) + T_{++}(a) {\cal O}(-b)\big) + (\hbox{\rm corrections})
\ee
Let's discuss the possible form of the corrections.  In two dimensions no corrections are necessary, by an argument from section \ref{sect:endpoint}.\footnote{In $d = 2$ take the correlator of (\ref{scalaroperatorid}) with an operator $X$.
If $X$ is in the conformal family of ${\cal O}$ the two sides agree by (\ref{GOexpansion}).  If $X$ is not in the conformal family of ${\cal O}$ both sides vanish by the conformal Ward identities \cite{Belavin:1984vu}.}  But in higher
dimensions corrections are necessary.  The left-hand side of (\ref{scalaroperatorid}) only involves the conformal family of ${\cal O}$.  But on the right-hand side the ${\cal O} \, T_{++}$ OPE can generate conformal families
with spin in addition to the desired conformal family of ${\cal O}$.\footnote{By the conformal Ward identities $\langle T_{\mu\nu} {\cal O}_i {\cal O}_j \rangle \sim \langle {\cal O}_i {\cal O}_j \rangle \sim \delta_{ij}$, so we don't need
to worry about generating other scalar conformal families \cite{Penedones:2016voo,Simmons-Duffin:2016gjk}.}  The correction terms should be a sum of products of operators at $a$ and $-b$, chosen to cancel the contribution of these spinning
conformal families.

Now we can use (\ref{deltaHA}) to compute the change in the modular Hamiltonian for subregion A.  For this we need the mirror operators
\beas
&& \widetilde{\partial_+^k T_{++}}(-b) = (-1)^k \partial_+^k T_{++}(b) \\
&& \widetilde{\partial_+^k {\cal O}}(-b) = (-1)^k \partial_+^k {\cal O}(b)
\eeas
Note that since CPT acts by $x^\pm \rightarrow - x^\pm$ it preserves the regulator (\ref{regulator}).  So we have
\bea
\delta H_A & = & {i \epsilon \over 2} {2 \theta^{{d \over 2} - 1} (2\sigma)^{d + 1} \over C_{{\cal OO}T}} \int_{-\infty}^\infty {ds \over 1 + \cosh s} \sum_{n = 0,2,4,\ldots} \beta_n \sigma^n (\partial_a - \partial_b)^n \\
\nonumber
& & \qquad \Big[{\cal O}(a)\big\vert_{s - i \pi} T_{++}(b) \big\vert_s - T_{++}(b) \big\vert_s {\cal O}(a) \big\vert_{s + i \pi} + T_{++}(a) \big\vert_{s-i\pi} {\cal O}(b)\big\vert_s - {\cal O}(b)\big\vert_s T_{++}(a)\big\vert_{s+i\pi}\Big] \\
\nonumber
& & + (\hbox{\rm corrections})
\eea
where the corrections cancel the contribution of any spinning conformal families to $\delta H_A$.  Thus to identify $\delta H_A$ as an operator it suffices to
consider a correlator with ${\cal O}(y)$.  This will also enable us to make sense of complex modular flow.  So we consider
\bea
&&\langle \delta H_A {\cal O}(y) \rangle = {i \epsilon \over 2} {2 \theta^{{d \over 2} - 1} (2\sigma)^{d + 1} \over C_{{\cal OO}T}} \int_{-\infty}^\infty {ds \over 1 + \cosh s} \sum_{n = 0,2,4,\ldots} \beta_n \sigma^n (\partial_a - \partial_b)^n \\
\nonumber
& & \qquad \langle\left({\cal O}(a)\big\vert_{s-ir} T_{++}(b) \big\vert_s - T_{++}(b) \big\vert_s {\cal O}(a) \big\vert_{s+ir} + T_{++}(a) \big\vert_{s-ir} {\cal O}(b)\big\vert_s - {\cal O}(b)\big\vert_s T_{++}(a)\big\vert_{s+ir}\right){\cal O}(y)\rangle
\eea
Our strategy is to start at $r = 0$, where the correlator is well-defined, and continue to $r = \pi$.  To make the starting point well-defined we take $y^2 > 0$ and $y^- > 0$ so that ${\cal O}(y)$ is spacelike separated
from the positive $x^+$ axis, which is where the other operators are located.  Modular flow acts by
\be
{\cal O}(x^+,x^-)\big\vert_s = {\cal O}(e^s x^+, e^{-s} x^-) \qquad T_{++}(x^+,x^-)\big\vert_s = e^{2s} T_{++}(e^s x^+,e^{-s}x^-)
\ee
which we'll assume makes sense even for complex $s$.

As $r \rightarrow \pi$ we can use the identity (\ref{GOexpansion}) to do the sum.  To see this it's convenient to rewrite everything in terms of
\be
\tilde{a}^- = e^{s - ir} a \big\vert_{r \rightarrow \pi} \qquad \tilde{a}^+ = e^{s + i r} a \big\vert_{r \rightarrow \pi} \qquad \tilde{b} = - e^s b \qquad \tilde{\theta} = e^{-2s}\theta
\ee
(the notation means we're keeping track of how $\tilde{a}^-$ and $\tilde{a}^+$ approach $-e^s a$).  This leads to\footnote{We're denoting the appropriate values of $a,\,b$ by placing subscripts on $G$.  The factor $-e^{-s}$ in (\ref{GOflow}) arises by combining the $e^{2s}$ from modular flow of $T_{++}$ with the fact that (from footnote \ref{footnote:odd}) $(2 \sigma)^{d+1}$ is odd under $\sigma \rightarrow - \sigma$, so in terms of tilded variables $\theta^{{d \over 2} - 1} (2 \sigma)^{d+1} =
- e^{-3s} \tilde{\theta}^{{d \over 2} - 1} (2 \tilde{\sigma})^{d+1}$.}
\be
\label{GOflow}
\langle \delta H_A {\cal O}(y) \rangle = {i \epsilon \over 2} \int_{-\infty}^\infty {ds \over 1 + \cosh s} \, \left(-e^{-s}\right) \big[\langle G_{\tilde{a}^-,\tilde{b}} {\cal O}(y) \rangle - \langle G_{\tilde{a}^+, \tilde{b}} {\cal O}(y) \rangle\big]
\ee
This relates $\delta H_A$ to $G_{\tilde{a}^\pm,\tilde{b}}$, so as promised $\delta H_A$ only involves the conformal family of ${\cal O}$.
To identify the specific operators that appear in $\delta H_A$ we proceed as follows.
Using (\ref{scalarGO}) to evaluate the correlators, and setting $z = e^s$, the $b$ dependence drops out and we're left with
\be
\label{zflow}
\langle \delta H_A {\cal O}(y) \rangle = - {i \epsilon \over (\Delta-1) y^-} \int_{0^+}^\infty {dz \over (z + 1)^2} {1 \over z} \left[{1 \over (e^{ir} a y^- z + y^2)^{\Delta - 1}} - {1 \over (e^{-ir} a y^- z + y^2)^{\Delta - 1}}\right]
\ee
The individual terms have poles at $z = 0$ and $z = -1$, and as $r \rightarrow \pi$ there's a cut along the positive real axis that starts at $z = {y^2 / a y^-}$.\footnote{Note that the pole at $z = 0$ cancels between the two terms in
(\ref{zflow}), so there's no harm in starting the integral at $0^+$.}  In the first term the cut hits the integration contour from above,
but in the second term it hits it from below, so we end up with a contour that wraps around the cut.

\centerline{\includegraphics[width=5cm]{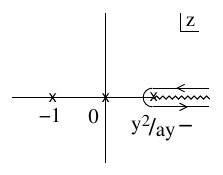}}

\noindent
To evaluate $\langle \delta H_A {\cal O}(y) \rangle$ it's convenient to deform the contour so it wraps around the poles instead.
\be
\langle \delta H_A {\cal O}(y) \rangle = - {i \epsilon \over (\Delta-1) y^-} \varointclockwise {dz \over (z + 1)^2} {1 \over z} {1 \over (-a y^- z + y^2)^{\Delta - 1}}
\ee
(The contour encircles the poles at $z = 0$ and $z = -1$ in a clockwise direction.)  To identify the endpoint contribution it suffices to set $a = 0$, in which case
\be
\langle \delta H_A {\cal O}(y) \rangle = - {i \epsilon \over (\Delta-1) y^-} \varointclockwise {dz \over (z + 1)^2} {1 \over z} \, {1 \over (y^2)^{\Delta - 1}}
\ee
But the integral vanishes, and since $\delta H_A$ is within the conformal family of ${\cal O}$ we reach the conclusion that for a scalar perturbation
\[
\delta H_A^{\rm endpoint} = 0
\]
Although this result by itself is not very exciting, in the next section these calculations will let us formulate a general conjecture for the endpoint contribution to $\delta H_A$.

\section{A general conjecture\label{sect:general}}
Given our explicit results for scalar and stress tensor perturbations we're in a position to conjecture a general result.  The steps are a bit abstract, so we illustrate them in the
concrete context of a 2-D CFT in appendix \ref{appendix:generaln}.

Consider a field $J^{(n)}$ that transforms with integer weight $n$ under modular flow.
\be
J^{(n)}(x^+,x^-,{\bf x}_\perp)\big\vert_s = e^{ns} J^{(n)}(e^s x^+,e^{-s} x^-,{\bf x}_\perp)
\ee
A scalar primary has weight $n = 0$ while $T_{++}$ has weight $n = 2$.  Suppose we perturb by
\be
G_{a,b} = \int_{-b}^a dx^+ J^{(n)}(x^+,0,0)
\ee
To proceed we could conjecture the existence of an operator identity analogous to (\ref{scalaroperatorid}).  But more directly, we can simply conjecture that the
appropriate generalization of (\ref{GOflow}) to this situation is
\be
\label{GJflows}
\langle \delta H_A J^{(n)}(y) \rangle = {i \epsilon \over 2} \int_{-\infty}^\infty {ds \over 1 + \cosh s} \, \left(-e^s\right)^{n-1} \big[\langle G_{\tilde{a}^-,\tilde{b}} J^{(n)}(y) \rangle - \langle G_{\tilde{a}^+, \tilde{b}} J^{(n)}(y) \rangle\big]
\ee
where
\be
\tilde{a}^\pm = e^{s \pm ir} a \big\vert_{r \rightarrow \pi} \qquad\quad \tilde{b} = - e^s b
\ee
This agrees with (\ref{GOflow}) when $n = 0$, and one can check that it agrees with (\ref{HATintegral}) when $n = 2$ and $d = 2$.  The factor $\left(-e^s\right)^{n-1}$ can be understood as arising from the fact that under modular flow
\be
G_{a,b} \big\vert_s = \int_{-b}^a dx^+ e^{ns} J^{(n)}(e^s x^+) = e^{(n-1)s} G_{a e^s,b e^s}
\ee
Continuing $s \rightarrow s \pm i \pi$ the prefactor becomes $\left(-e^s\right)^{n-1}$.

Assuming this conjecture is correct, by changing variables to $z = e^s$ we have
\be
\label{GJflowz}
\langle \delta H_A J^{(n)}(y) \rangle = i \epsilon \int_{0^+}^\infty {dz \over (z+1)^2} \, (-z)^{n-1} \big[\langle G_{\tilde{a}^-,\tilde{b}} J^{(n)}(y) \rangle - \langle G_{\tilde{a}^+, \tilde{b}} J^{(n)}(y) \rangle\big]
\ee
where we're starting the integral at $z = 0^+$, corresponding to a cutoff at large negative modular time, and where
\be
\tilde{a}^\pm = e^{\pm ir} a z \big\vert_{r \rightarrow \pi} \qquad\quad \tilde{b} = - b z
\ee
By examining some special cases we expect that $\langle G_{\tilde{a}^\pm,\tilde{b}} J^{(n)}(y) \rangle$ has a cut along the positive $z$ axis that starts at $z = y^2 / a y^-$, with the behavior\footnote{Here $J^{(n)}$ is taken to be a primary operator of dimension $\Delta$ in a CFT.  The result is easy to
see in two dimensions, where $\langle J^{(n)}(x) J^{(n)}(0) \rangle \sim {1 \over \left(x^+\right)^{\Delta + n} \left(x^-\right)^{\Delta - n}}$.  In general dimensions, besides scalars (\ref{scalarGO}), another instructive case
is a spin-1 primary with $\langle j_\mu(x) j_\nu(0) \rangle = \left(g_{\mu\nu} - {2 x_\mu x_\nu \over x^2}\right) {1 \over \left(x^2\right)^\Delta}$.}
\be
\label{largez}
\langle G_{\tilde{a}^\pm,\tilde{b}} J^{(n)}(y) \rangle \sim \left\lbrace\begin{array}{ll}
z & \hbox{\rm as $z \rightarrow 0$} \\
{1 \over z^{\Delta + n - 1}} & \hbox{\rm as $z \rightarrow \infty$}
\end{array}\right.
\ee
So besides the cut the individual terms have poles at $z = -1$ and (for $n < 0$) at $z = 0$.  The pole at $z = 0$ is avoided by cutting off the integral at large negative modular time.

In (\ref{GJflowz}) the $b$ dependence cancels and can be discarded by sending $b \rightarrow \infty$.  We're left with a contour that wraps around the branch cut, but thanks to the large $z$ behavior in (\ref{largez})
the contour can be deformed to encircle the poles instead.  We can strip off the spectator operator $J^{(n)}(y)$, and to isolate the endpoint contribution it suffices to set $a = 0$.  Then we're left with
\be
\delta H_A^{\rm endpoint} = i \epsilon \varointclockwise {dz \over (z + 1)^2} (-z)^{n-1} \int_0^\infty dx^+ J^{(n)}(x^+)
\ee
Evaluating the integral we find that
\be
\delta H_A^{\rm endpoint} = \left\lbrace
\begin{array}{ll}
-2 \pi \epsilon (n - 1) \int_0^\infty dx^+ J^{(n)}(x^+) \quad & \hbox{\rm for $n = 2,3,4,\ldots$} \\[5pt]
\quad 0 & \hbox{\rm for $n = \ldots,-1,0,1$}
\end{array}\right.
\ee

It's curious that endpoint contributions only arise for $n \geq 2$.  For this we offer the intuitive explanation that endpoint contributions arise when the generator $G$ can move operators from $A$ to $\bar{A}$ or vice versa along the $x^+$ Rindler horizon.  This property is explicit in the case of stress-tensor perturbation in two dimensions, as we mentioned below \eqref{ConformalDeltaH}, and fits with the appearance of endpoint contributions in modular Hamiltonians under shape deformations \cite{Faulkner:2016mzt}.
To probe this take a scalar primary ${\cal O}(x^+,x^-,{\bf x}_\perp)$ and insert it on the boundary between $A$ and $\bar{A}$, which means setting $x^+ = 0$. 
The OPE can generate derivatives of ${\cal O}$,
\be
J^{(n)}(0) {\cal O}(x) \sim \cdots + B^{(n) \, \lambda} \partial_\lambda O(x) + \cdots
\ee
Could this OPE include a $\partial_+ {\cal O}$, which could move the operator between $A$ and $\bar{A}$ along the $x^+$ Rindler horizon?  We can test this by looking at the modular weights of both sides.
The OPE coefficient $B^{(n) \, \lambda}$ is built from the metric and $x^\mu$.  The metric has modular weight zero, while $x^\mu = (x^+,x^-,{\bf x}_\perp)$ has modular weights $(-1,+1,0)$.  So when $x^+ = 0$ the modular weight of
$B^{(n) \, \lambda}$ is either zero or positive.  And since $\partial_+ {\cal O}$ has modular weight 1, we see that $\partial_+ {\cal O}$ can only appear in the OPE with an operator of weight $n \geq 1$.  In
appendix \ref{appendix:OPE} we carry out a more refined analysis and show that, although $\partial_+ {\cal O}$ indeed appears in the OPE for $n \geq 1$, it can only appear in the equal-time commutator $[J^{(n)}(0), {\cal O}(x)]$ for $n \geq 2$.
So it takes an operator with modular weight $n \geq 2$ to move fields between $A$ and $\bar{A}$ along the $x^+$ Rindler horizon.  This generalizes the familiar fact that in two dimensions it takes $T_{++}$, an operator with modular
weight $2$, to generate a reparametrization of $x^+$.

We conclude by noting that more generally we could perturb the state by
\be
G = \int dx^+ d^{d-2}x_\perp f(x^+,{\bf x}_\perp) J^{(n)}(x^+,0,{\bf x}_\perp)
\ee
Then the endpoint contribution to $\delta H_A$ would be a superposition
\be
\delta H_A^{\rm endpoint} = \left\lbrace
\begin{array}{ll}
- 2 \pi \epsilon (n-1) \int d^{d-2} x_\perp \, f(0,{\bf x}_\perp) \int_0^\infty dx^+ J^{(n)}(x^+,0,{\bf x}_\perp) \quad & \hbox{\rm for $n = 2,3,4,\ldots$} \\[5pt]
\quad 0 & \hbox{\rm for $n = \ldots,-1,0,1$}
\end{array}\right.
\ee
This can be combined with the analogous result for $\delta H_{\bar{A}}$ to obtain the endpoint contribution to the extended modular Hamiltonian given
in the introduction.\footnote{The easiest way to obtain $\delta H_{\bar{A}}$ is to apply a CPT transformation, use the result for $\delta H_A$, then transform back.}
\be
\delta \Hext^{\rm endpoint} = \left\lbrace
\begin{array}{ll}
- 2 \pi \epsilon (n-1) \int d^{d-2} x_\perp \, f(0,{\bf x}_\perp) \int_{-\infty}^\infty dx^+ J^{(n)}(x^+,0,{\bf x}_\perp) \quad & \hbox{\rm for $n = 2,3,4,\ldots$} \\[5pt]
\quad 0 & \hbox{\rm for $n = \ldots,-1,0,1$}
\end{array}\right.
\ee
This result clearly relies on vacuum modular flow \cite{Bisognano:1976za} but doesn't seem to require conformal invariance.  So we believe it should hold in a general quantum field theory.

\section{Conclusions\label{sect:conclusions}}
Given a state $\vert \Psi \rangle$ and a division of space into $A \cup \bar{A}$, the extended modular Hamiltonian $\Hext$ is a well-defined operator, free from short-distance ambiguities.  However we've seen that
$\Hext$ is sensitive to detailed properties of the state at the boundary between $A$ and $\bar{A}$.  We explored this for states that are small perturbations of the vacuum, where to first order we identified an endpoint
contribution (\ref{deltaHext}) which is present in $\delta \Hext$.  Endpoint contributions only arise for perturbations generated by operators of modular weight 2 or greater.  Intuitively we believe this is because
only such operators can move degrees of freedom between $A$ and $\bar{A}$ along the $x^+$ Rindler horizon.

There are several directions that call for further development.  For example, we determined the endpoint contributions to first order for small perturbations about the vacuum.  How are endpoint contributions modified at higher orders in perturbation theory?  Also our explicit calculations in sections \ref{sect:endpoint} and \ref{sect:scalar} relied on conformal invariance to constrain the behavior of correlation functions.  But the general result conjectured
in section \ref{sect:general} doesn't seem to require conformal invariance.  Instead it only appears to rely on the universal features of vacuum modular flow \cite{Bisognano:1976za}.  Can the general conjecture be tested and
verified in non-conformal theories?

It would be interesting to make contact between the results in this paper and related studies that have appeared in the literature. This includes work on modular Hamiltonians in excited states \cite{Lashkari:2015dia,Sarosi:2016oks,Casini:2017roe,Sarosi:2017rsq,Lashkari:2018oke,Arias:2020qpg}
and for shape-deformed regions using path integral methods \cite{Allais:2014ata,Faulkner:2016mzt,Lewkowycz:2018sgn,Balakrishnan:2020lbp}. When the perturbing operator is the stress tensor, preliminary work with path integral methods gives an endpoint contribution identical to the one derived here and in \cite{Faulkner:2016mzt}. We hope to report a detailed study on this in the near future.  It would also be interesting to make contact with the work on 
entanglement entropy for excited states that appeared in \cite{Rosenhaus:2014woa,Rosenhaus:2014zza,Speranza:2016jwt,Belin:2018juv,Belin:2019mlt}.

\bigskip
\goodbreak
\centerline{\bf Acknowledgements}
\noindent
We thank Bartek Czech, Lampros Lamprou and Shubho Roy for discussions. DK is supported by U.S.\ National Science Foundation grant PHY-1820734.  GL is supported in part by the Israel Science Foundation under grant 447/17.  PN acknowledges support from
Israel Science Foundation grant 447/17 for the work in sections \ref{sect:conformal} and \ref{sect:perturbation} and
U.S.\ National Science Foundation grant PHY-1820734 for the work in sections \ref{sect:endpoint}, \ref{sect:scalar} and \ref{sect:general}. DS thanks IAS, Tsinghua University for support and hospitality during the initial stages of this project.

\appendix
\section{Modular Hamiltonians for conformally-excited states\label{appendix:excited}}
For completeness we review the derivation \cite{Das:2018ojl} of the modular Hamiltonian for a certain class of excited states, namely states that can be obtained from the vacuum by applying a conformal transformation.  Start with the vacuum modular Hamiltonian for an interval $(w_1,w_2)$.
\be
\label{H0w}
\Hext^{(0,w)}_{(w_1,w_2)} = 2 \pi \int_{-\infty}^\infty dw \, {(w_2 - w)(w - w_1) \over w_2 - w_1} \, T_{++}(w) + (\hbox{\rm right-movers})
\ee
The notation $\Hext^{(0,w)}$ emphasizes that the CFT is in its ground state in the $w$ coordinates.  We'll suppress the right-moving contribution in what follows.
So although $w$ is a spatial coordinate, on the $t = 0$ slice it can be identified with the light-front coordinate $w^+$.

Suppose there's a conformal map $g : z \rightarrow w$ that takes an excited state in the $z$ coordinates to the ground state in $w$.

\centerline{\includegraphics{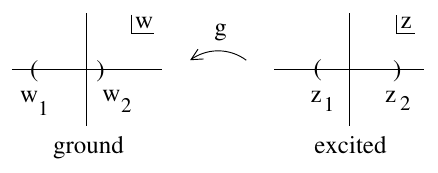}}

The excited state modular Hamiltonian $\Hext_{(z_1,z_2)}$ is the pull-back of the vacuum modular Hamiltonian $\Hext^{(0,w)}_{(w_1,w_2)}$.  So in (\ref{H0w}) we set
$w = g(z)$ with $w_1 = g(z_1)$, $w_2 = g(z_2)$ and $T(w) = \left({dz \over dw}\right)^2 T(z) + ({\rm c-number})$.  This gives the excited state modular Hamiltonian for the interval $(z_1,z_2)$.
\be
\label{Hz}
\Hext_{(z_1,z_2)} = 2 \pi \int_{-\infty}^\infty dz \, {(g(z_2) - g(z)) (g(z) - g(z_1)) \over g'(z)(g(z_2) - g(z_1))} \, T_{++}(z)
\ee
Relabeling $z_1 \rightarrow u$, $z_2 \rightarrow v$ gives (\ref{Hmodg}).
Note that in the $z$ coordinates the vacuum modular Hamiltonian for the same interval $(z_1,z_2)$ would be\footnote{We're distinguishing between $\Hext^{(0,w)}_{(w_1,w_2)}$
and $\Hext^{(0,z)}_{(z_1,z_2)}$ since -- except for $g \in SL(2,{\mathbb R})$ as mentioned in the next paragraph -- they're not related by the map $g$.}
\be
\label{H0z}
\Hext^{(0,z)}_{(z_1,z_2)} = 2 \pi \int_{-\infty}^\infty dz \, {(z_2 - z)(z - z_1) \over z_2 - z_1} \, T_{++}(z)
\ee
Relabeling $z_1 \rightarrow u$, $z_2 \rightarrow v$ gives (\ref{Hmod0}).

As one check of these results, recall that the vacuum state is invariant under global conformal transformations.  So when the map $g : z \rightarrow {a z + b \over c z + d}$ is an $SL(2,{\mathbb R})$
transformation we should have $\Hext_{(z_1,z_2)} = \Hext^{(0,z)}_{(z_1,z_2)}$.  It's straightforward to check that this is indeed the case.

It's worth noting that, as an integral of the stress tensor, the modular Hamiltonian (\ref{Hz}) generates an infinitesimal conformal transformation $z^+ \rightarrow z^+ + \delta z^+$ with
\be
\label{deltaz+}
\delta z^+ = 2 \pi \, {(g(z_2) - g(z)) (g(z) - g(z_1)) \over g'(z)(g(z_2) - g(z_1))}
\ee
So for excited states that can be obtained from the vacuum by a conformal transformation, modular flow retains its local geometric character.  It is simply flow along the conformal Killing
vector (\ref{deltaz+}).  Note that, as expected, the endpoints of the interval ($z = z_1$ and $z = z_2$) are fixed points of the flow.

\section{Perturbation expansion for $\delta H_A$\label{appendix:perturbation}}
For completeness we give a formal derivation of (\ref{deltaHA}), summarizing results in the literature.

Imagine we have a separable\footnote{meaning with a countable orthonormal basis.} Hilbert space that can be factored, ${\cal H} = {\cal H}_A \otimes {\cal H}_{\bar{A}}$.
By going to a Schmidt basis we can take the ground or reference state $\vert 0 \rangle$ to have a thermofield form in terms of orthonormal states $\vert i \rangle$.  
\be
\vert 0 \rangle = {1 \over \sqrt{Z}} \sum_i e^{-\beta E_i / 2} \, \vert i \rangle \otimes \vert i \rangle
\ee
We'll assume the reduced density matrices
\bea
\label{rho0A}
&& \rho^{(0)}_A = {\rm Tr}_{\bar{A}} \, \vert 0 \rangle \langle 0 \vert = {1 \over Z} \sum_i e^{-\beta E_i}  \vert i \rangle \langle i \vert \\
\nonumber
&& \rho^{(0)}_{\bar{A}} = {\rm Tr}_{A} \, \vert 0 \rangle \langle 0 \vert = {1 \over Z} \sum_i e^{-\beta E_i}  \vert i \rangle \langle i \vert
\eea
have maximal rank -- an assumption which should be safe in field theory.  

If we perturb the state $\vert \psi \rangle = e^{-i \epsilon G} \vert 0 \rangle$ then the reduced density matrix for region $A$ has matrix elements\footnote{To save writing we denote $\vert ij \rangle = \vert i \rangle_A \otimes \vert j \rangle_{\bar{A}}$,
$\langle ij \vert = \langle i \vert_A \otimes \langle j \vert_{\bar{A}}$.}
\be
\langle l \vert \rho_A \vert m \rangle = {1 \over Z} \sum_{i,j,k} e^{-\beta(E_i + E_j)/2} \langle lk \vert e^{-i \epsilon G} \vert ii \rangle \langle jj \vert e^{i \epsilon G} \vert mk \rangle
\ee
We expand to first order in $\epsilon$ and assume $G = G_A \otimes G_{\bar{A}}$.  Then the first-order change in the reduced density matrix is
\bea
\nonumber
\langle l \vert \delta \rho_A \vert m \rangle & = & {1 \over Z} \sum_i e^{-\beta (E_i + E_m)/2}  (-i \epsilon) \langle l \vert G_A \vert i \rangle \langle m \vert G_{\bar{A}} \vert i \rangle \\
& & \! + {1 \over Z} \sum_i e^{-\beta (E_i + E_l)/2}  (i \epsilon) \langle i \vert G_A \vert m \rangle \langle i \vert G_{\bar{A}} \vert l \rangle
\eea
We define mirror operators $\widetilde{G}_{\bar{A}}$ by their matrix elements
\be
\label{mirror}
{}_A \langle i \vert \widetilde{G}_{\bar{A}} \vert j \rangle_A = {}_{\bar{A}} \langle j \vert G_{\bar{A}} \vert i \rangle_{\bar{A}}
\ee
Note that $\widetilde{G}_{\bar{A}}$ is defined on the Hilbert space ${\cal H}_A$, even though $G_{\bar{A}}$ acts on the Hilbert space for the complement ${\cal H}_{\bar{A}}$.
This definition lets us write
\bea
\nonumber
\langle l \vert \delta \rho_A \vert m \rangle & = & {1 \over Z} \sum_i e^{-\beta (E_i + E_m)/2}  (-i \epsilon) \langle l \vert G_A \vert i \rangle \langle i \vert \widetilde{G}_{\bar{A}} \vert m \rangle \\
& & \! + {1 \over Z} \sum_i e^{-\beta (E_i + E_l)/2}  (i \epsilon) \langle l \vert \widetilde{G}_{\bar{A}} \vert i \rangle \langle i \vert G_A \vert m \rangle
\eea
In terms of the unperturbed density matrix (\ref{rho0A}) this means
\be
\delta \rho_A = - i \epsilon G_A (\rho^{(0)}_A)^{1/2} \widetilde{G}_{\bar{A}} (\rho^{(0)}_A)^{1/2} + i \epsilon (\rho^{(0)}_A)^{1/2} \widetilde{G}_{\bar{A}} (\rho^{(0)}_A)^{1/2} G_A
\ee
To get an expression for $\delta H_A$ we use the expansion of the logarithm developed in \cite{Sarosi:2017rsq,Lashkari:2018tjh}, which to first order reads
\be
H_A^{(0)} + \delta H_A = - \log \left(\rho_A^{(0)} + \delta \rho_A\right) = - \log \rho_A^{(0)} - {1 \over 2} \int_{-\infty}^\infty {ds \over 1 + \cosh s} \, (\rho^{(0)}_A)^{-{1 \over 2} - {i s \over 2 \pi}} \,
\delta \rho_A \, (\rho^{(0)}_A)^{-{1 \over 2} + {i s \over 2 \pi}}
\ee
Given our expression for $\delta \rho_A$ this means
\be
\delta H_A = {i \epsilon \over 2}  \int_{-\infty}^\infty {ds \over 1 + \cosh s} \left( (\rho^{(0)}_A)^{-{1 \over 2} - {i s \over 2 \pi}} G_A (\rho^{(0)}_A)^{1/2} \widetilde{G}_{\bar{A}} (\rho^{(0)}_A)^{i s \over 2 \pi} - (\rho^{(0)}_A)^{- {i s \over 2 \pi}} \widetilde{G}_{\bar{A}} (\rho^{(0)}_A)^{1/2} G_A (\rho^{(0)}_A)^{-{1 \over 2} + {i s \over 2 \pi}} \right)
\ee
It's convenient to define modular-flowed operators
\be
{\cal O} \big\vert_s = \Delta^{-{i s \over 2 \pi}} {\cal O} \Delta^{i s \over 2 \pi} = (\rho^{(0)}_A)^{- {i s \over 2 \pi}} {\cal O} (\rho^{(0)}_A)^{i s \over 2 \pi}
\ee
(the second equality holds for operators that just act on ${\cal H}_A$) and write the result in the form given in (\ref{deltaHA}).
\be
\delta H_A = {i \epsilon \over 2}  \int_{-\infty}^\infty {ds \over 1 + \cosh s} \left( G_A \big\vert_{s-i\pi} \widetilde{G}_{\bar{A}} \big\vert_s - \widetilde{G}_{\bar{A}} \big\vert_s G_A \big\vert_{s+i\pi} \right)
\ee

We conclude by noting that the mirror operators defined in (\ref{mirror}) can be identified with the mirror operators introduced in \cite{Papadodimas:2012aq,Papadodimas:2013jku}, which for a division into half-spaces means they can be obtained
by CPT conjugation.  To see this we set
\be
G_{\bar{A}} \vert j \rangle_{\bar{A}} = \sum_i G_{\bar{A}ij} \vert i \rangle_{\bar{A}}
\ee
so that matrix elements are denoted
\be
{}_{\bar{A}} \langle i \vert G_{\bar{A}} \vert j \rangle_{\bar{A}} = G_{\bar{A}ij}
\ee
Then from the definition (\ref{mirror}) we have
\be
\label{eqn1}
{}_A \langle i \vert \widetilde{G}_{\bar{A}} \vert j \rangle_A = {}_{\bar{A}} \langle j \vert G_{\bar{A}} \vert i \rangle_{\bar{A}} = G_{\bar{A}ji} = G^*_{\bar{A}ij}
\ee
where the last equality assumes that $G_{\bar{A}}$ is Hermitian.  Now consider the anti-unitary operator $J$ of Tomita-Takesaki theory, which acts on the Schmidt basis
by $J (\vert i \rangle_A \otimes \vert j \rangle_{\bar{A}}) = \vert j \rangle_A \otimes \vert i \rangle_{\bar{A}}$, and note that
\be
\label{eqn2}
\langle il \vert J G_{\bar{A}} J \vert jm \rangle = \langle il \vert J G_{\bar{A}} \vert mj \rangle = \langle il \vert J \sum_k G_{\bar{A}kj} \vert mk \rangle = \langle il \vert \sum_k G^*_{\bar{A}kj} \vert km \rangle = G^*_{\bar{A}ij} \delta_{lm}
\ee
where the next-to-last equality uses anti-linearity.  Comparing (\ref{eqn1}) and (\ref{eqn2}), we can identify $\widetilde{G}_{\bar{A}} = J G_{\bar{A}} J$.  For a division into half-spaces modular conjugation is the same as CPT conjugation, $J = {\sf CPT}$.\footnote{See \cite{Bisognano:1976za} and section 5 of \cite{Witten:2018lha}, where the operation is denoted ${\sf CRT}$.}  Thus for a division into half-spaces $\widetilde{G}_{\bar{A}}$ and $G_{\bar{A}}$ are CPT conjugates.

\section{General $n$ in $d = 2$\label{appendix:generaln}}
In this appendix we illustrate the steps outlined in section \ref{sect:general} in the concrete setting of a 2-D CFT.
Consider a primary operator $J^{(n)}(x^+,x^-)$ of modular weight $n \in {\mathbb Z}$ and conformal dimension $\Delta$, normalized so that
\begin{equation}
\langle J^{(n)}(x) \, J^{(n)}(0) \rangle = {1 \over (x^+)^{\Delta+n} (x^-)^{\Delta - n}}
\end{equation}
That is, $J^{(n)}$ is a primary of dimension $({\Delta + n \over 2},\,{\Delta - n \over 2})$.
Then $G_{a,b} = \int_{-b}^a dx^+ J^{(n)}(x^+,0)$ satisfies (taking $y^+ < -b$ to avoid singularities)
\begin{equation}
\langle G_{a,b} \, J^{(n)}(y) \rangle = - {1 \over \Delta + n - 1} \left({1 \over (a - y^+)^{\Delta + n - 1}} - {1 \over (-b-y^+)^{\Delta + n - 1}}\right) {1 \over (-y^-)^{\Delta - n}}
\end{equation}
and the general conjecture (\ref{GJflowz}) becomes (with $r$ approaching $\pi$ from below)
\begin{equation}
\label{2dflow}
\langle \delta H_A J^{(n)}(y) \rangle = {i \epsilon \over \Delta + n - 1}{1 \over (-y^-)^{\Delta - n}} \int_{0^+}^\infty {dz \over (z+1)^2} (-z)^{n-1} \left[{1 \over (e^{ir} a z - y^+)^{\Delta + n - 1}}
- {1 \over (e^{-ir} a z - y^+)^{\Delta + n - 1}}\right]
\end{equation}
Note that the $b$ dependence cancels.  The two terms can be combined into a single integration contour that wraps around the branch cut as shown below (\ref{zflow}).  For $n \geq 1$ the
contour can be deformed to encircle the pole at $z = -1$, giving
\begin{eqnarray}
\label{HJcontour}
\langle \delta H_A J^{(n)}(y) \rangle & = & {i \epsilon \over \Delta + n - 1} {1 \over (-y^-)^{\Delta - n}} \varointclockwise {dz \over (z+1)^2} (-z)^{n-1} {1 \over (- a z - y^+)^{\Delta + n - 1}} \\[5pt]
\nonumber
& = & {2 \pi \epsilon \over \Delta + n - 1} {1 \over (-y^-)^{\Delta - n}} {(n-1) y^+ + \Delta a \over (a - y^+)^{\Delta + n}}
\end{eqnarray}
As $a \rightarrow 0$ we have
\begin{eqnarray}
\nonumber
\langle \delta H_A J^{(n)}(y) \rangle & = & - 2 \pi \epsilon (n - 1) {1 \over \Delta + n - 1} {1 \over (-y^+)^{\Delta + n - 1} (-y^-)^{\Delta - n}} \\[5pt]
& = & - 2 \pi \epsilon (n -1) \langle G_{\infty,0} \, J^{(n)}(y) \rangle
\end{eqnarray}
So we identify
\begin{equation}
\delta H_A = - 2 \pi \epsilon (n-1) G_{\infty,0} = - 2 \pi \epsilon (n-1) \int_0^\infty dx^+ J^{(n)}(x^+,0) \qquad n = 1,2,3,\ldots
\end{equation}
If $n \leq 0$ the contour in (\ref{HJcontour}) encircles the poles at $z = 0$ and $z = -1$ and the integral vanishes as $a \rightarrow 0$.
Note that although the light-ray operator $\int_{-\infty}^\infty dx^+ J^{(n)}(x^+,0)$ has a vanishing correlator with $J^{(n)}(y)$ and indeed annihilates the vacuum \cite{Kravchuk:2018htv},
the half-light-ray operator $G_{\infty,0}$ has a non-trivial correlator.

To establish (\ref{2dflow}) in this setting one might ask for an operator identity analogous to (\ref{T++operatorid}).  The relevant 3-point function is (with normalization fixed by the conformal Ward identity)
\be
\langle T(z) J^{(n)}(w_1) J^{(n)}(w_2) \rangle = {(\Delta + n) / 2 \over (z^+ - w_1^+)^2 (z^+ - w_2^+)^2 (w_1^+ - w_2^+)^{\Delta + n - 2} (w_1^- - w_2^-)^{\Delta - n}}
\ee
and the analog of (\ref{GTexpansion}) is
\bea
\label{GJexpansion}
\nonumber
&& \langle G_{a,b} \, J^{(n)}(y) \rangle = {(2 \sigma)^3 \over \Delta + n} \sum_{k=0,2,4,\ldots} \gamma_k \sigma^k (\partial_a - \partial_b)^k \langle \big[J^{(n)}(a,0) T(-b) + T(a) J^{(n)}(-b,0)\big] J^{(n)}(y) \rangle \\[5pt]
&& \gamma_0 = 1, \qquad \gamma_2 = - {(\Delta + n - 3)(\Delta + n - 8) \over 3 (\Delta+n) (\Delta+n+1)}, \qquad \ldots
\eea
This can be seen by expanding both sides in powers of ${\sigma \over y^+ - \delta}$.  The operator identity follows by stripping off the spectator $J^{(n)}(y)$.

\section{OPEs and commutators\label{appendix:OPE}}
In this appendix we study the commutator $[J^{(n)},{\cal O}]$, with a view toward understanding when $J^{(n)}$ is able to move degrees of freedom from $A$ to $\bar{A}$ or vice versa. 
To get oriented it's useful to record some explicit OPE formulas.  For the OPE of a spin-1 primary $J_\mu$ with a scalar primary ${\cal O}_1$ we have ($s^\mu = x^\mu - y^\mu$ and $s = \sqrt{s^2}$)\footnote{We can derive this by considering the OPE limit of a three-point function involving a spin-1 and two scalar primaries. See e.g. \cite{Simmons-Duffin:2016gjk}.}
\bea
& & J_\mu(x) {\cal O}_1(y) \stackrel{x \rightarrow y}{\sim} {c_{J{\cal O}_1{\cal O}_2} \over s^{\Delta_J + \Delta_1 - \Delta_2}} \left(A_\mu(s) {\cal O}_2(y) + B_\mu{}^\lambda(s) \partial_\lambda {\cal O}_2(y) + \cdots \right) \\
\nonumber
& & A_\mu(s) = {s_\mu \over s} \\
\nonumber
& & B_\mu{}^\lambda(s) = {s \over 2 \Delta_2} \delta_\mu{}^\lambda + {\Delta_J + \Delta_2 - \Delta_1 - 1 \over 2 \Delta_2} {s_\mu s^\lambda \over s}
\eea
Here $c_{J{\cal O}_1{\cal O}_2}$ is the coefficient in the three-point function $\langle J_\mu {\cal O}_1 {\cal O}_2 \rangle$.
For the stress tensor the analogous result is \cite{Osborn:1993cr}
\bea
&& T_{\mu\nu}(x) {\cal O}(y) \stackrel{x \rightarrow y}{\sim} {1 \over s^d} \left(A_{\mu\nu}(s) {\cal O}(y) + B_{\mu\nu}{}^\lambda(s) \partial_\lambda {\cal O}(y) + \cdots \right) \\
\nonumber
&&A_{\mu\nu}(s) = {\rm const.} \left({s_\mu s_\nu \over s^2} - {1 \over d} g_{\mu\nu}\right) \\
\nonumber
&&B_{\mu\nu}{}^\lambda(s) = {\rm const.} \left(s_\mu \delta_\nu{}^\lambda + s_\nu \delta_\mu{}^\lambda - g_{\mu\nu} s^\lambda + (d-2) {s_\mu s_\nu s^\lambda \over s^2}\right)
\eea
The general pattern is a singular function of $s^2$ times factors of $s_\mu$ and the metric.

Let's begin by working on a fixed-time hypersurface $t = 0$ and dividing space into $A = \lbrace x > 0 \rbrace$ and $\bar{A} = \lbrace x < 0 \rbrace$.  Insert a probe operator ${\cal O}$ on the surface $x = 0$ (the boundary that separates
$A$ from $\bar{A}$).  We want to see if the equal-time commutator
$[J^{(n)},{\cal O}]$ can generate a term $\sim \partial_x {\cal O}$ that could move the operator across the boundary.

Recall that the commutator can be obtained from the OPE by taking the difference of two $i \epsilon$ prescriptions.\footnote{This may be more familiar in the context of Green's functions,
where the commutator Green's function is the difference of two Wightman functions \cite{Birrell:1982ix}.}  Note that for
\be
s^2 = -(t \pm i \epsilon)^2 + x^2 + \vert {\bf x}_\perp \vert^2
\ee
the two $i \epsilon$ prescriptions are the same at equal times.  So a function of $s^2$ cannot give rise to a non-trivial equal-time commutator.  The only way to get a non-vanishing commutator is to have the
explicit appearance of $t$ with a free vector index in the OPE, which will give a $\pm i \epsilon$ in the numerator.  Also note that since $x = 0$, we don't need to worry about the combination $s^\lambda \partial_\lambda {\cal O}$
generating $\partial_x {\cal O}$.  This means $J^{(n)}$ must also have a free $x$ index.  Thus it takes an operator with spin 2 such as $T_{tx}$ to produce an equal-time commutator $\sim \partial_x {\cal O}$.

We're more interested in commutators on a null plane, which we define by infinitely boosting a fixed-time hypersurface.  Decomposing $T_{tx}$ into operators of definite modular weight, only the component with highest modular
weight $T_{++}$ survives the boost.  Thus it takes an operator with modular weight at least 2 to generate a light-front commutator $\sim \partial_+ {\cal O}$, which we interpret as moving the operator from one region to
the other along the $x^+$ Rindler horizon.  More generally we expect an operator of modular weight $n$ to produce a commutator $\sim \partial_+^{n-1} {\cal O}$
for $n > 0$.

\section{CFT conventions\label{appendix:conventions}}
To make contact with the CFT conventions of \cite{Ginsparg:1988ui} consider a free boson with action, stress tensor and correlator
\bea
\nonumber
&& S = - {A \over 2} \int d^2x \sqrt{-h} \, h^{\alpha\beta} \partial_\alpha \phi \partial_\beta \phi \\
&& T_{++} = - {2 \over \sqrt{-h}} {\delta S \over \delta h^{++}} = A \partial_+ \phi \partial_+ \phi \\
\nonumber
&& \langle \phi(x) \phi(y) \rangle = - {1 \over 4 \pi A} \log \left(-(x^+-y^+)(x^--y^-)\right)
\eea
Then with a Wightman $i \epsilon$ prescription
\be
\label{dphidphi}
\langle \partial_+ \phi(x) \partial_+ \phi(y) \rangle = - {1 \over 4 \pi A} {1 \over (x^+-y^+-i\epsilon)^2}
\ee
Comparing to (2.16) in \cite{Ginsparg:1988ui} fixes the field normalization $A = 1/\pi$.  Then to agree with the stress tensor convention (2.17) in \cite{Ginsparg:1988ui} we set
\be
T(x) = - 2 \pi T_{++}(x)
\ee
Incidentally the 2-point function of a Hermitian operator should be positive at coincident points.  This is at least formally true in (\ref{dphidphi}) thanks to the $i \epsilon$ prescription.


\providecommand{\href}[2]{#2}\begingroup\raggedright\endgroup

\end{document}